\documentstyle[prl,aps,epsfig,multicol,amssymb,amsfonts]{revtex}
\tighten
\renewcommand{\narrowtext} 
{\begin{multicols}{2}\global\columnwidth20.5pc} 
\renewcommand{\widetext}
{\end{multicols}\global\columnwidth42.5pc} 
\input psfig

\begin{document} 
\draft 
\author{} 

\narrowtext
{\bf \noindent Comment on ``Analytic Structure of One-Dimensional Localization
     Theory: Re-Examining Mott's Law''\\}

In a recent Letter A. O. Gogolin \cite{Gogolin00} has challenged the
established point of view that Mott's prediction for the dynamical
conductivity of a localized electron system is correct. The 
intuitive argument \cite{Mott67} leads in one dimension to a
dynamical conductivity of the form $\omega^2\ln^2\omega$. Later, the
precise, asymptotical  result  
\begin{equation} \label{eq:1}
\Re\ \frac{\sigma(\omega)}{\sigma_0} = 
\nu^2 \big( \ln^2 \nu - \frac{\pi^2}{4} + (2 {\cal C}-3)\ln \nu - C + \cdots \big)
\end{equation}
( $\nu=2\omega\tau$, ${\cal C}$ denotes the Euler-Mascheroni constant
$0.5772\ldots$ and $\sigma_0 = e^2 v_F
\tau/\pi$ per spin).
has been derived by several authors using different methods, see e.~g. Refs.
\onlinecite{Berezinskii73,Abrikosov78,Pastur82}. (We are not aware of
any analytical prediction for the constant $C$.)
Gogolin presents a purely formal calculation which yields
\begin{equation} \label{eq:2}
\Re\ \frac{\sigma(\omega)}{\sigma_0} = \frac{1}{3}\  \nu ^2 \big( \ln ^3 \frac{1}{\nu} + \cdots\big)
\end{equation}
(eq. (22) in Ref. \onlinecite{Gogolin00} with  $\sigma_0=4$).
In view of the mentioned variety of works corroborating
Mott's conclusion this is quite unexpected. If Gogolin were right,
then one of the thought to be most profound chapters in
localization theory would have to be rewritten. In fact, however,
as we will demonstrate below, he is not. 

Gogolin's analysis starts from the famous recursion
equations derived first by Berezinskii\cite{Berezinskii73}.
The equations can be solved in a standard manner by mapping
them to a differential equation. Gogolin's claim is that
the previous solution of this equation is incorrect and
hence also the conductivity law $\omega^2\ln^2 \omega$
derived thereof. He argues that previous authors have not properly
taken into account discreteness of the spectrum
of the equation.

A simple method to check Berezinskii's result
is to solve the recursion equations for the conductivity numerically.
(For details see Ref. \cite{Gogolin78}.)
The algorithm is very stable and has been used down to frequencies
$\nu = 5\cdot10^{-6}$ where $M=10^8$ in a calculation with 40 digits
(fixed) precision. For even larger $M=2\cdot 10^8$
or more digits, e.g. 60, $\sigma$ does not change implying that
rounding errors are irrelevant.
Fig. 1 shows our result. The agreement of the numerical data
with the Mott/Berezinskii-solution is perfect over more than 3 decades
while the data is completely incompatible with Gogolin's $\ln^3\omega$ term. 

One may ask where Gogolin's approach fails.
We believe that the problem stems from the
``leading logarithmic approximation'',
the only step in the calculation which is not exact.
The expression eq. (19) in Gogolin's paper derived within this
approximation may be 
\begin{figure}
\includegraphics[width=0.8\columnwidth,clip]{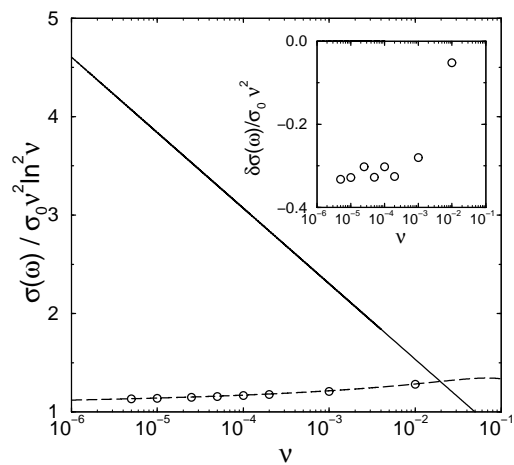}
\caption{Dynamical conductivity from solving the
Berezinskii recursion equations ($\nu=2\omega\tau$).
Numerical solution ($\circ$), Berezinskii's solution, eq. (\ref{eq:1}) (dashed),
Gogolin's result, eq. (\ref{eq:2}) (solid). Inset: Determining $C$ by
subtracting first three terms in eq. (\ref{eq:1}) from numerical data:
$C\approx 0.3$.} 
\label{rmfres} 
\end{figure}
\noindent sufficient for
obtaining the leading term $\propto i \omega $. However, the real
part of $\sigma$ is of higher order in $\omega$ and presumably 
to this order corrections exist that have been ignored by Gogolin
and that cancel the $\ln^3\omega$ term. We also mention, that in
contrast to Gogolin's statements the length $\ell \ln(1/\nu)$
has been identified and discussed in the literature as a relevant
scale, e.~g. in Ref. \cite{Gorkov83}.

We thank A. Mildenberger, A.~D. Mirlin, D.~G. Polyakov, L. Schweitzer
and P. W\"olfle  for stimulating discussions. Support from the SFB 195
der DFG is gratefully acknowledged.\\

\noindent Ferdinand Evers \\
\indent Institut f\"ur Nanotechnologie, \\
\indent Forschungszentrum Karlsruhe, \\
\indent D-76021 Karlsruhe, Germany \\
\\
\noindent Berndt M. Gammel \\
\indent Infineon Technologies AG, \\
\indent D-81609  M\"unchen, Germany

\pacs{\mbox{\hspace{-2cm} PACS numbers: 72.15.Rn, 73.20.Fz}}

\end{multicols}
\end{document}